\documentclass[12pt]{iopart}
\usepackage[dvips]{graphicx}
\usepackage[latin1]{inputenc}
\usepackage{bbm}
\usepackage{psfrag}
\usepackage{iopams}

\newcommand{\beq}{\begin{equation}} 
\newcommand{\eeq}{\end{equation}}
\newcommand{\bea}{\begin{eqnarray}} 
\newcommand{\eea}{\end{eqnarray}}

\begin{document} 
 
\title{Solvable model of a self-gravitating system}

\author{Lapo Casetti and Cesare Nardini}
\address{Dipartimento di Fisica e Astronomia and 
Centro per lo Studio delle Dinamiche Complesse (CSDC), 
Universit\`a di Firenze, and Istituto Nazionale di Fisica Nucleare (INFN), Sezione di Firenze, 
via G.~Sansone 1, 50019 Sesto Fiorentino (FI), Italy}
\ead{\mailto{lapo.casetti@unifi.it},\mailto{cesare.nardini@gmail.com}}

 
\begin{abstract} 
We introduce and discuss an effective model of a self-gravitating system whose equilibrium thermodynamics can be solved in both the microcanonical and the canonical ensemble, up to a maximization with respect to a single variable. Such a model can be derived from a model of self-gravitating particles confined on a ring, referred to as the self-gravitating ring (SGR) model, allowing a quantitative comparison between the thermodynamics of the two models. Despite the rather crude approximations involved in its derivation, the effective model compares quite well with the SGR model. Moreover, we discuss the relation between the effective model presented here and another model introduced by Thirring forty years ago. The two models are very similar and can be considered as examples of a class of minimal models of self-gravitating systems. 
\end{abstract} 
 
 

\date{\today} 
 
\section{Introduction}

Systems of classical particles mutually interacting via gravitational forces can model the behavior of many objects in the universe (globular clusters, elliptical galaxies, clusters of galaxies) as long as other interactions are negligible compared to the gravitational ones \cite{BinneyTremaine}. When the number $N$ of particles is large the direct numerical simulation of such systems is a heavy task \cite{Nbody} and it would be reasonable and useful to approach them via equilibrium statistical mechanics. However, self-gravitating systems do not have a ``true'' thermal equilibrium for two main reasons \cite{IspolatovCohen}: $(i)$ the gravitational potential is singular for vanishing distance between two particles, making (at least part of) the system collapse in states with infinite density and $(ii)$ particles that do not collapse tend to escape the system (evaporation). From a physical point of view the first problem can be easily solved. No real system exists where the only non-negligible interaction is classical gravity at {\em all} length scales: either the interacting ``particles'' are macroscopic bodies like stars or galaxies, or quantum effects must be taken into account below a certain length scale. In both cases, a length scale exists (the size of the bodies or the scale where quantum effects set in) below which the potential has no longer the classical gravitational form. If one is not interested in small-scale details, the potential can be regularized by replacing it with a softened one at short distances \cite{softening,fermions} or by directly considering self-gravitating fermions \cite{fermions}. To solve the second problem is less straightforward and one is forced to (somehow artificially) confine the particles in a finite volume. However, on physical grounds such an approximation is reasonable since in many cases the evaporation rate is slow compared to the other time scales in the system \cite{FanelliMerafinaRuffo}. A regularized and confined self-gravitating system has a thermal equilibrium in both the canonical and the microcanonical ensemble \cite{Kiessling}. Such a system can thus be studied within the framework of equilibrium statistical mechanics, although its behaviour in the two ensembles is very different: it can be considered a prototype of systems with ensemble inequivalence \cite{CampaDauxoisRuffo}, showing e.g.\ a core-halo phase with negative specific heat in the microcanonical ensemble which is replaced by a discontinuous phase transition from a clustered to a gas phase in the canonical ensemble \cite{DeVega}.

Canonical and microcanonical MonteCarlo simulations of a full self-gravitating system in three
spatial dimensions are heavy \cite{DeVega}, although they may be a little easier than the direct
integration of the equations of motion. This suggests to look for simplified models which may be
easier to study; one main simplification comes from considering models which are effectively
one-dimensional. Many such models have been introduced in the last decades, ranging from the sheet
model \cite{HohlFeix} and the shell model \cite{YoungkinsMiller} to the self-gravitating ring (SGR)
model \cite{SGR1}. The latter is particularly interesting because the interaction among the
particles is given by the full three-dimensional gravitational potential (regularized at small
distances), while the particles are confined on a ring. This yields a behaviour that is
qualitatively very close to that found in three-dimensional systems, although allowing a much
simpler study. In the limit of an infinite number of particles, the model can be studied in the
mean-field approximation with a very efficient numerical technique \cite{SGR2}, showing that in the
microcanonical ensemble there is a phase transition separating a homogeneous high-energy phase from
a clustered phase. An independent analytical argument supporting the existence of such a transition
has been given in \cite{gravring}.

The aim of the present paper is to introduce and discuss an effective model which approximates the SGR model and is exactly solvable (up to a maximization with respect to a single variable which has to be performed numerically). Although the approximations involved are rather crude, the behaviour of this effective model is very close to that of the SGR model, not only at a qualitative level but also at a semi-quantitative one: this suggests that the approximations made do capture most of the important physics. 

The paper is organized as follows. After recalling the main features of the SGR model we introduce and discuss our effective model, and we present its solution in both the microcanonical and the canonical ensembles; the details of the calculations are reported in two appendices. Then, we compare our model with one introduced by Thirring forty years ago \cite{Thirring}: the two models are indeed very similar, although the latter was not aimed at approximating any particular explicit model. We end with some remarks and a discussion of some open issues. 

\section{Model}

We now want to introduce and discuss our model. Before doing that, we briefly recall the main features of the self-gravitating ring (SGR) model.

\subsection{Self-gravitating ring model}

The SGR model describes $N$ identical classical particles of mass $m$ constrained to move on a ring of radius $R$ and mutually interacting via a regularized gravitational potential. Its Hamiltonian is \cite{SGR1}
\beq
\label{hamiltonianasgr1}
{\cal H}_{\mathrm{SGR}}=\frac{1}{2mR^2}\sum _{i=1}^{N}P_{i}^2- \frac{1}{2}\sum _{i,j = 1}^{N}\frac{Gm^2}{\sqrt{2}R\sqrt{1-\cos \left(\vartheta _{i}-\vartheta _{j}\right)+\alpha }}\,,
\eeq
where $\vartheta_i \in (-\pi,\pi]$ is the angular coordinate of the $i$-th particle,  
\beq
P_i = mR^2\frac{d\vartheta_i}{dt}
\eeq
is its angular momentum, $R$ is the radius of the circle and $G$ is Newton's gravitational constant. The constant $\alpha > 0$ provides the short-distance regularization: when $\vartheta_i - \vartheta_j \ll 1$ the interaction between the $i$-th and the $j$-th particle is effectively harmonic on a length scale of the order of 
\beq
d_\alpha = R\sqrt{2\alpha}~.
\eeq 
The SGR model can thus describe a self-gravitating system when $d_\alpha \ll R$, i.e., $\alpha \ll 1$. On the other hand, in the opposite limit $\alpha \to \infty$ the SGR model becomes equivalent to the ferromagnetic hamiltonian mean-field (HMF) model \cite{HMF}, that is a model of fully coupled planar classical spins with ferromagnetic interactions.

In the microcanonical ensemble, the fundamental quantity to compute is the density of states. For the Hamiltonian (\ref{hamiltonianasgr1}) it can be written as
\beq
\omega _{N} (E)=\frac{1}{N!}\int_{-\infty}^{\infty} dP_{1} \cdots dP_{N}\int_{-\pi}^\pi d\vartheta_{1} \cdots d\vartheta_{N}\,\delta \left({\cal H}_{\mathrm{SGR}}-E\right)~.
\label{omegadim}
\eeq
Defining a characteristic time scale $\tau$ as 
\beq
\tau = \sqrt{\frac{R^3}{GNm}}
\eeq
and dimensionless momenta as 
\beq
p_i = \frac{d\vartheta}{d\tau}~, ~~~i = 1,\ldots,N
\eeq
one can write a dimensionless Hamiltonian $\tilde \mathcal{H}$ as
\beq
\label{H}
\tilde \mathcal{H}=\frac{1}{2}\sum _{i=1}^{N}p_{i}^2- \frac{1}{2N\sqrt{2}}\sum_{i,j=1}^{N}\frac{1}{\sqrt{1-\cos \left(\vartheta _{i}-\vartheta _{j}\right)+\alpha }}\,.
\eeq
Since 
\beq
\tilde \mathcal{H} = \frac{\tau^2}{mR^2} {\cal H}_{\mathrm{SGR}}~, 
\eeq
the density of states (\ref{omegadim}) can be written as 
\beq
\omega _{N} (E)=\frac{m^N R^{2N}}{\tau^N N!}\int_{-\infty}^{\infty} dp_{1} \cdots dp_{N}\int_{-\pi}^\pi d\vartheta_{1} \cdots d\vartheta_{N}\,\delta \left[\frac{m R^2}{\tau^2}\left(\tilde \mathcal{H}-\tilde{E}\right)\right]\,,
\label{omegadim2}
\eeq
where $\tilde{E} = \frac{\tau^2}{mR^2} E$, 
and using the properties of the $\delta$ function it becomes
\beq
\omega _{N} (\tilde{E})=\frac{m^{N-1} R^{2(N-1)}}{\tau^{N-2} N!} \tilde{\omega}_{N} (\tilde{E})
\eeq
where $\tilde{\omega}_{N}$ is the dimensionless density of states
\beq
\tilde{\omega}_{N} (\tilde{E}) = \int_{-\infty}^{\infty} dp_{1} \cdots dp_{N}\int_{-\pi}^\pi d\vartheta_{1} \cdots d\vartheta_{N}\,\delta \left(\tilde \mathcal{H}-\tilde{E}\right)~.
\label{omega-adim} 
\eeq
Before going on, let us note that the entropy derived from the full density of states (\ref{omegadim2}) is extensive in the thermodynamic limit $N\to\infty$, $R\to\infty$, $\frac{R}{N} \to \mathrm{const.}$; moreover, the adimensional Hamiltonian (\ref{H}) is also extensive, due to the $\frac{1}{N}$ rescaling of the coupling between the particles. The Kac prescription for making extensive a long-range interaction \cite{CampaDauxoisRuffo} is here obtained via a suitable choice of the adimensionalization procedure.

From now on we will consider only the dimensionless Hamiltonian (\ref{H}) and density of states (\ref{omega-adim}). For ease of notation, we will drop the tilda's and simply denote them by $\mathcal{H}$ and $\omega_N(E)$, where also $E$ is adimensional.

\subsection{Effective model}

We now want to approximate the Hamiltonian of the SGR model, or more precisely the adimensional Hamiltonian (\ref{H}), in order to make it soluble, i.e., such that the density of states (\ref{omega-adim}) can be computed. 

Numerical simulations of the dynamics of the SGR model reported in \cite{SGR1} have shown that, at a given energy, particles can be roughly divided into three classes according to their dynamical behaviour: {\em cluster} particles, {\em gas} particles, and {\em halo} particles. Cluster particles are tightly bound in a cluster and never get far from it; gas particles move almost freely around the ring; halo particles have a complicated dynamics that is somehow intermediate between the other two. The relative population of particles in the three classes depends on the energy (or temperature): at low energy almost all the particles are cluster particles, while at high energy all the particles are gas particles. 

The strategy we are going to implement in order to define an effective model is to consider only the first two classes of particles (cluster and gas) and to assume that each particle belongs to one of the two classes. This allows a simplification of the potential energy which makes the model soluble.

Let us then assume that $N_g$ particles, with $1 \leq N_g \leq N$, are gas particles. Then, writing the Hamiltonian (\ref{H}) as 
\beq
{\cal H} = \frac{1}{2}\sum_{i=1}^N p_i^2 + V(\vartheta_1,\ldots,\vartheta_N),
\eeq
we can split the potential energy $V$ into three parts:
\bea
V(\vartheta_1,\ldots,\vartheta_N) & = & V_{\mathrm{gas}}(\vartheta_1,\ldots,\vartheta_{N_g}) + V_{\mathrm{cluster}}(\vartheta_{N_g + 1},\ldots,\vartheta_N) \nonumber \\
& + & V_{\mathrm{int}}(\vartheta_1,\ldots,\vartheta_N)~,
\eea
where
\begin{eqnarray}
V_{\mathrm{gas}}(\vartheta_1,\ldots,\vartheta_{N_g}) & = & - \frac{1}{2N\sqrt{2}}\sum_{i,j=1}^{N_g} v\left(\vartheta _{i}-\vartheta _{j}\right)\,, \label{Vgas1}\\
V_{\mathrm{cluster}}(\vartheta_{N_g + 1},\ldots,\vartheta_{N}) & = & - \frac{1}{2N\sqrt{2}}\sum_{i,j=N_g + 1}^{N}v\left(\vartheta _{i}-\vartheta _{j}\right)\,, \label{Vcluster1}\\
V_{\mathrm{int}}(\vartheta_1,\ldots,\vartheta_{N}) & = & - \frac{1}{N\sqrt{2}}\sum_{i=1}^{N_g}\sum_{j = N_g +1}^{N}v\left(\vartheta _{i}-\vartheta _{j}\right)\,, \label{Vint1}
\end{eqnarray}
where
\beq
v(x) = \frac{1}{\sqrt{1-\cos x +\alpha }}~. 
\eeq
Up to this point we have only rewritten the potential energy in a different form. However, this form naturally allows to introduce the approximations which make the model soluble. Let us now discuss the approximations.
\begin{description}
\item[$(i)$] Since the $N_g$ gas particles are essentially free particles, as far as their interaction $V_{\mathrm{gas}}$ is concerned we consider them as uniformly distributed on the circle. The interaction energy (\ref{Vgas1}) between these particles is then a constant:  
\beq
V_{\mathrm{gas}} = - \gamma \frac{N_g^2}{2N\sqrt{2}}
\label{Vgas}
\eeq
where
\beq
\gamma = \frac{1}{2\pi} \int_{-\pi}^\pi v(x)\, dx = \frac{2}{\pi\sqrt{2 + \alpha}} \mathcal{K}\left(\frac{2}{2 + \alpha}\right)~
\label{gamma}
\eeq
and ${\cal K}(x)$ is the complete elliptic integral of the first kind. 
\item[$(ii)$] We consider the $N - N_g$ remaining particles as confined in a cluster. We assume the cluster is tight, i.e., the particles are all close to each other:
\beq
\vartheta_i - \vartheta_j \ll 1 ~~\,\, \forall i,j = N_g +1,\ldots,N~.
\label{assum}
\eeq
We can thus expand the interaction energy (\ref{Vcluster1}) among these particles up to the harmonic order, and write
\beq
V_{\mathrm{cluster}} = - \frac{\left(N - N_g\right)^2}{2N\sqrt{2\alpha}} + \frac{1}{8\alpha N \sqrt{2\alpha}} \sum_{i,j=N_g + 1}^{N}\left(\vartheta_i -  \vartheta_j \right)^2~;
\label{Vcluster}
\eeq
such an approximation is reliable if an assumption stronger than (\ref{assum}) holds, i.e.,
\beq
\frac{\left(\vartheta_i - \vartheta_j\right)^2}{\alpha} \ll 1 ~~\,\, \forall i,j = N_g +1,\ldots,N~.
\label{assum2}
\eeq
Moreover, since the particles in the cluster do not ``feel'' the $\mathbb{S}^1$ topology of the circle, we assume that
\beq
-\infty < \vartheta_i < +\infty, ~~ \forall i = N_g +1,\ldots,N~;
\eeq
this will allow the analytical computation of the configurational integrals in the density of states. 
\item[$(iii)$] As far as the interaction (\ref{Vint1}) between cluster and gas particles is concerned, we note that as long as the assumption (\ref{assum}) holds, the typical distance between a gas particle and a cluster particel is much larger than typical interparticle distances in the cluster, so that we may assume that all the cluster particles are in the same location, i.e., $\vartheta = 0$. Being the gas particles uniformly distributed on the circle, this yields a constant for $V_{\mathrm{int}}$, i.e.,
\beq
V_{\mathrm{int}} = - \gamma \frac{N_g(N - N_g)}{N\sqrt{2}} ~,
\label{Vint}
\eeq
where $\gamma$ is given by (\ref{gamma}).
\end{description}
The Hamiltonian of our effective model is then
\beq
{\cal H}_{\mathrm{eff}} = \frac{1}{2}\sum_{i = 1}^N p_i^2 + V_{\mathrm{eff}}~,
\label{Heff}
\eeq
where
\beq
V_{\mathrm{eff}} = - V_0(N,N_g,\alpha) + \frac{\mu}{2} \sum_{i,j=N_g + 1}^{N} \left(\vartheta_i -  \vartheta_j \right)^2~, 
\eeq
and where we have set 
\beq
V_0(N,N_g,\alpha) = \frac{\left(N - N_g\right)^2}{2N\sqrt{2\alpha}} + \gamma\left[ \frac{N_g(N - N_g)}{N\sqrt{2}} + \frac{N_g^2}{2N\sqrt{2}} \right]~
\eeq
and
\beq
\mu = \frac{1}{2N (2\alpha)^{3/2}}~.
\eeq

\section{Microcanonical and canonical thermodynamics}

Let us now discuss the solution of the effective model in the microcanonical and canonical ensembles. In the limit $N \to \infty$, at fixed $N_g$ the model is exactly solvable in both ensembles. However, $N_g$ is not {\em a priori} assigned and must be fixed in a self-consistent way. The simplest way to do so is to take into account all the possible values of $N_g$; as we shall show in the following, in the limit $N \to \infty$ the model is still solvable up to a maximization (resp. minimization) in a single variable which must be performed numerically, whose physical meaning is just to determine the value of $N_g$ that maximizes the entropy (resp. minimizes the free energy). 

\subsection{Microcanonical ensemble}

To solve the model in the microcanonical ensemble we need to calculate the 
entropy density
\beq
s(\varepsilon) = \lim_{N\to\infty} \frac{1}{N} \log \omega_N(\varepsilon)~, 
\label{sdef}
\eeq
where we have introduced the energy density $\varepsilon =  \frac{E}{N}$ and $\omega_N$ is the density of states (\ref{omega-adim}), where $\cal H$ is replaced by $\mathcal{H}_\mathrm{eff}$:
\begin{eqnarray}
{\omega}_{N} (\varepsilon) & = & \int_{-\infty}^{\infty} dp_{1} \cdots dp_{N}\int_{-\pi}^\pi d\vartheta_{1} \cdots d\vartheta_{N_g} \int_{-\infty}^\infty d\vartheta_{N_g} \cdots d\vartheta_{N}\,\delta \left({\cal H}_{\mathrm{eff}}-{N\varepsilon}\right) \nonumber \\
& = & \sum_{N_g = 0}^N \frac{N!}{N_g!\left(N - N_g \right)!} \int_{-\infty}^{\infty} dp_{1} \cdots dp_{N}\int_{-\pi}^\pi d\vartheta_{1} \cdots d\vartheta_{N_g}  \label{omegaeff}\\
& \times & \int_{-\infty}^\infty d\vartheta_{N_g+1} \cdots d\vartheta_{N}\,\delta \left[\frac{1}{2}\sum_{i = 1}^N p_i^2 + \frac{\mu}{2}\sum_{i,j=N_g + 1}^{N}\left(\vartheta_i -  \vartheta_j \right)^2 - V_0 -{N\varepsilon}\right] ~. \nonumber 
\end{eqnarray}
In the above expression we have summed over all the possible choices of $N_g$, properly counted by the degeneracy factor ${N \choose{N_g}}$. 

The calculation of the above integral is straightforward, albeit a bit involved, and can be performed following a procedure similar to that used in \cite{KastnerSchnetz,CasettiKastner}. The details are reported in \ref{app_micro}. It turns out that the entropy density in the thermodynamic limit is given by
\beq
s(\varepsilon) = \sup_{n_g \in [0,n_g^{\mathrm{max}}(\varepsilon)]} s(\varepsilon,n_g) ~.
\label{smax}
\eeq
where we have introduced the fraction of gas particles $n_g = \frac{N_g}{N}$ and $n_g^{\mathrm{max}}(\varepsilon)$ is the maximum fraction of gas particles allowed at a given energy density $\varepsilon$, given by Eq.\ (\ref{ngmax_app}). The explicit expression of $s(\varepsilon,n_g)$ is
\begin{eqnarray}
s(\varepsilon,n_g) & = &\frac{1-n_g}{2}\log\left[\frac{4\pi(2\alpha)^{3/2}}{(1-n_g)(2-n_g)} \right] + \frac{1}{2} \log \left( \frac{2\pi\sqrt{2}}{2-n_g}\right) \nonumber \\
& + & \frac{2-n_g}{2} \left[1 + \log a(n_g,\alpha,\varepsilon) \right] + n_g \log (2\pi) \label{s(ng)}\\
& - & n_g \log n_g - (1 - n_g) \log (1 - n_g) ~, \nonumber
\end{eqnarray}
with
\beq
a(n_g,\alpha,\varepsilon) = \frac{\gamma}{2\sqrt{2}} n_g(2-n_g) + \frac{\left(1-n_g\right)^2}{2\sqrt{2\alpha}} + \varepsilon~;
\eeq
one can check that $a(n_g,\alpha,\varepsilon)>0$ if $n_g \in [0,n_g^{\mathrm{max}}(\varepsilon)]$ and $\varepsilon > \varepsilon_{\mathrm{min}}$, where $\varepsilon_{\mathrm{min}} = - \frac{1}{2\sqrt{2\alpha}}$ is the absolute minimum of the potential energy per degree of freedom.

As anticipated, the solution of the effective model in the microcanonical ensemble amounts to finding the value $\overline{n_g}(\varepsilon)$ of $n_g$ realizing the extremum in (\ref{smax}). This can be easily done numerically, since the explicit form (\ref{s(ng)}) of $s(\varepsilon,n_g)$ is available. 

\subsubsection{Results for the thermodynamic quantities}

\begin{figure}
\center
\psfrag{k}{$\overline{n_g}$}
\psfrag{e}{$\varepsilon$}
\psfrag{0.8}{0.8}
\psfrag{0.6}{0.6}
\psfrag{0.4}{0.4}
\psfrag{0.2}{0.2}
\psfrag{1}{1}
\psfrag{-1}{-1}
\psfrag{-2}{-2}
\psfrag{-3}{-3}
\includegraphics[width=12cm,clip=true]{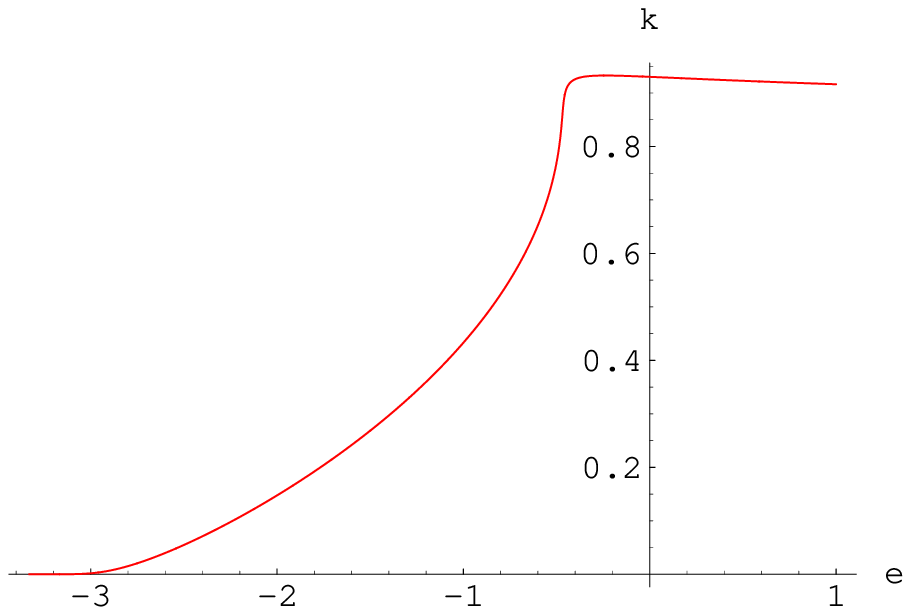}
\vspace{0.1cm}
\psfrag{T}{$T$}
\psfrag{e}{$\varepsilon$}
\psfrag{0.5}{0.5}
\psfrag{1}{1}
\psfrag{1.5}{1.5}
\psfrag{2}{2}
\psfrag{2.5}{2.5}
\psfrag{3}{3}
\psfrag{1}{1}
\psfrag{-1}{-1}
\psfrag{-2}{-2}
\psfrag{-3}{-3}
\includegraphics[width=12cm,clip=true]{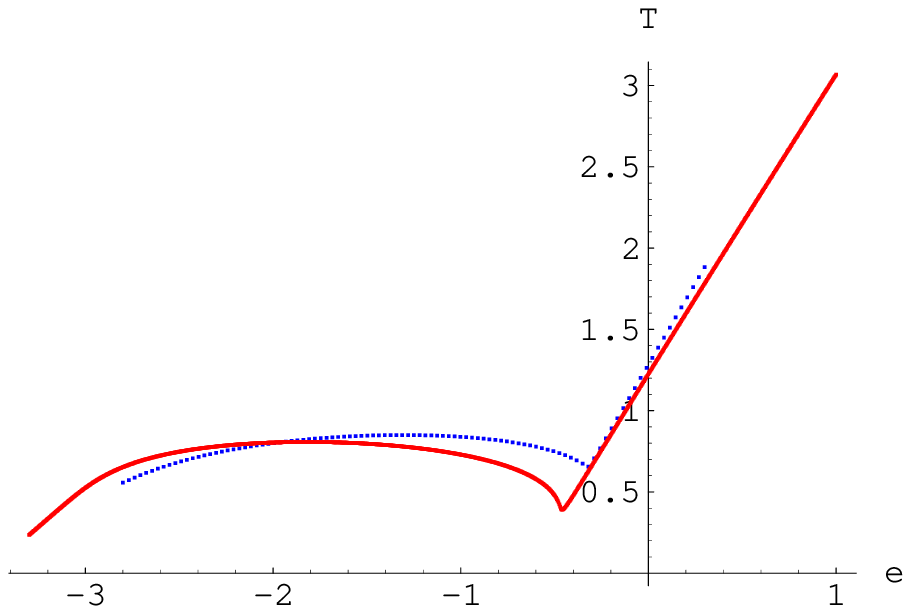}
\caption{Effective model in the microcanonical ensemble, $\alpha = 10^{-2}$. (top) Fraction of gas particles $\overline{n_g}(\varepsilon)$ (red line); (bottom) temperature $T(\varepsilon)$ computed for the effective model (red line) and for the SGR model (blue symbols).}
\label{fig_low}
\end{figure}

\begin{figure}
\center
\psfrag{k}{$\overline{n_g}$}
\psfrag{e}{$\varepsilon$}
\psfrag{0.8}{0.8}
\psfrag{0.6}{0.6}
\psfrag{0.4}{0.4}
\psfrag{0.2}{0.2}
\psfrag{1}{1}
\psfrag{20}{20}
\psfrag{-20}{-20}
\psfrag{-40}{-40}
\psfrag{-60}{-60}
\includegraphics[width=12cm,clip=true]{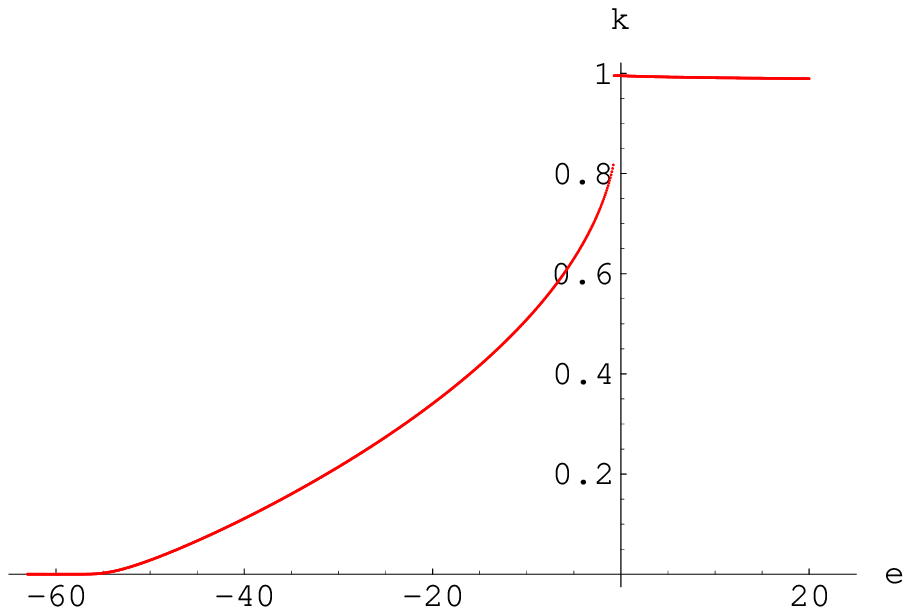}
\vspace{0.1cm}
\psfrag{T}{$T$}
\psfrag{e}{$\varepsilon$}
\psfrag{5}{5}
\psfrag{10}{10}
\psfrag{15}{15}
\psfrag{20}{20}
\psfrag{20}{20}
\psfrag{-20}{-20}
\psfrag{-40}{-40}
\psfrag{-60}{-60}
\includegraphics[width=12cm,clip=true]{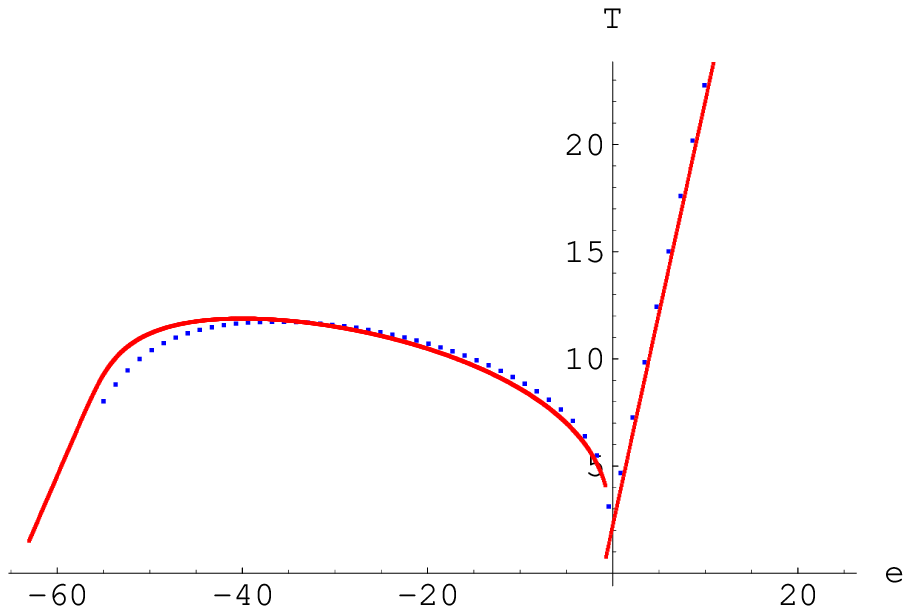}
\caption{As in Fig.\ \protect\ref{fig_low} for $\alpha = 3\times 10^{-5}$.}
\label{fig_high}
\end{figure}

In the following we report the results for the fraction of gas particles $\overline{n_g}(\varepsilon)$ and for the caloric curve, i.e., the temperature $T(\varepsilon) = \left(\frac{ds}{d\varepsilon}\right)^{-1}$, as a function of $\varepsilon$. We also compare the latter quantity with that computed for the SGR model via the numerical method introduced in \cite{SGR2}; for $\overline{n_g}(\varepsilon)$ such a comparison is impossible, because no such quantity is easily defined for the SGR model. In Fig.\ \ref{fig_low} we report $\overline{n_g}(\varepsilon)$ and $T(\varepsilon)$ computed for a softening parameter $\alpha = 10^{-2}$, as well as a comparison with $T(\varepsilon)$ for the SGR model; in Fig.\ \ref{fig_high} we report the same quantities for $\alpha = 3\times 10^{-5}$. 
The agreement with the SGR model is reasonably good already at $\alpha = 10^{-2}$ and becomes very good at $\alpha = 3\times 10^{-5}$. In both cases we find a phase transition from a homogeneous phase (characterized by $\overline{n_g}\simeq 1$) to a clustered phase while lowering $\varepsilon$ below a critical value $\varepsilon_c$; the critical energy is $\varepsilon_c\simeq -0.46$ for $\alpha = 10^{-2}$ and $\varepsilon_c\simeq -0.8$ for $\alpha = 3\times 10^{-5}$. These values should be compared with those found for the SGR model, i.e., $\varepsilon_c\simeq -0.32$ for $\alpha = 10^{-2}$ and $\varepsilon_c\simeq -0.5$ for $\alpha = 3\times 10^{-5}$. The agreement is good, especially for the lower value of $\alpha$. In the case  $\alpha = 10^{-2}$ the phase transition is continuous, while it is discontinuous (the temperature $T$ jumps between two different values at $\varepsilon_c$) at $\alpha = 3\times 10^{-5}$. We find indeed a microcanonical tricritical point which is located at $\alpha \simeq 5\times 10^{-3}$; in the case of the SGR model this point is located at $\alpha \simeq 10^{-4}$, so that also in this respect the two models are very similar. We stress that the above results come from a numerical maximization so that we have no rigorous proof of the existence of true singularities in the microcanonical ensemble for the effective model. However, the same holds for the SGR model too, as long as the softening parameter $\alpha$ is finite.

\begin{figure}
\center
\psfrag{a}{$\frac{d\overline{n_g}}{d\varepsilon}$}
\psfrag{e}{$\varepsilon$}
\psfrag{0.04}{0.04}
\psfrag{0.03}{0.03}
\psfrag{0.02}{0.02}
\psfrag{0.01}{0.01}
\psfrag{20}{20}
\psfrag{-20}{-20}
\psfrag{-40}{-40}
\psfrag{-60}{-60}
\includegraphics[width=12cm,clip=true]{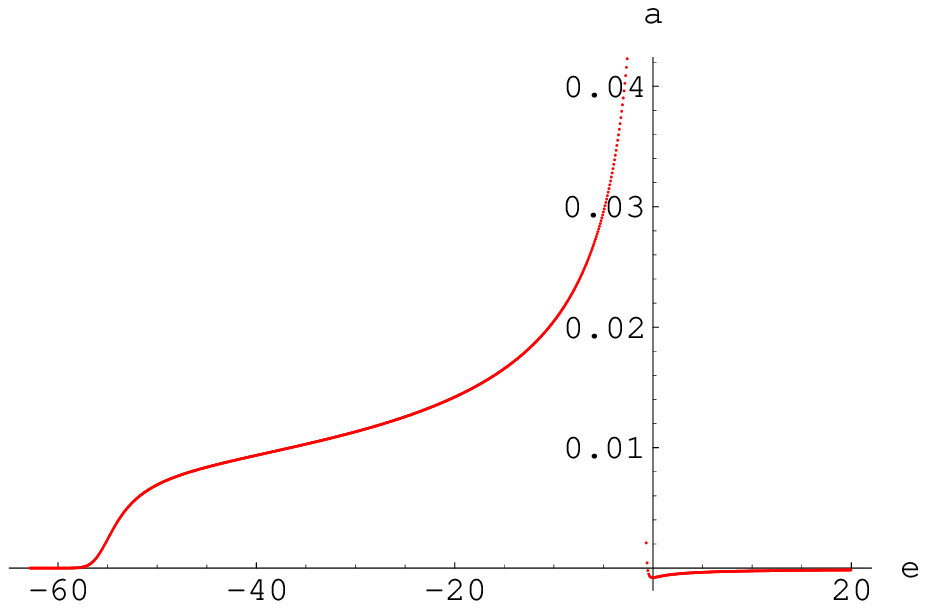}
\vspace{0.1cm}
\psfrag{a}{$\frac{d^2\overline{n_g}}{d\varepsilon^2}$}
\psfrag{e}{$\varepsilon$}
\psfrag{0.002}{0.002}
\psfrag{0.0015}{0.0015}
\psfrag{0.001}{0.001}
\psfrag{0.0005}{0.0005}
\psfrag{-0.0005}{-0.0005}
\psfrag{-0.001}{-0.001}
\psfrag{20}{20}
\psfrag{20}{20}
\psfrag{-20}{-20}
\psfrag{-40}{-40}
\psfrag{-60}{-60}
\includegraphics[width=12cm,clip=true]{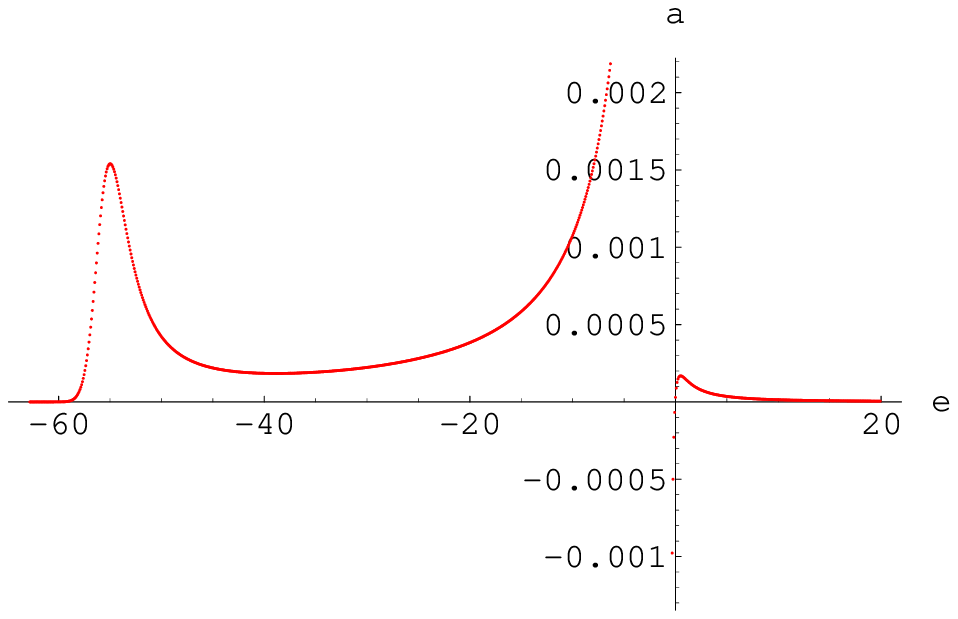}
\caption{Effective model in the microcanonical ensemble, $\alpha = 3\times 10^{-5}$. (top) First derivative of the fraction of gas particles,  $\frac{d\overline{n_g}}{d\varepsilon}$; (bottom) Second derivative $\frac{d^2\overline{n_g}}{d\varepsilon^2}$.}
\label{fig_dng}
\end{figure}

In the SGR model, the high-energy phase ($\varepsilon > \varepsilon_c$) is a perfect gas phase, as shown by the numerical calculations\footnote{In the limit $\alpha \to \infty$, when the SGR model becomes equivalent to the ferromagnetic HMF model, this can be shown analytically; however, for large values of $\alpha$ our effective model is not a good approximation.} reported in \cite{SGR2}. We should thus expect $\overline{n_g} \equiv 1$ for $\varepsilon > \varepsilon_c$. This does not happen, although $\overline{n_g}$ is very close to 1 ($\overline{n_g} \simeq 0.93$ for $\varepsilon \gtrsim \varepsilon_c$ when $\alpha = 10^{-2}$ and $\overline{n_g} \simeq 0.995$ for $\varepsilon \gtrsim \varepsilon_c$ when $\alpha = 3\times10^{-5}$). This is due to the presence of the degeneracy term ${N \choose{N_g}}$ in the density of states, which makes the entropy vanish for $n_g =1$ and $n_g = 0$, so that the extremum of the entropy (\ref{s(ng)}) can never be realized in the boundaries of the domain of $n_g$ when the energy density is strictly larger than its absolute minimum $\varepsilon_{\mathrm{min}}$. For the latter value of $\varepsilon$, however, we have $\overline{n_g} \equiv 0$, and there is a whole region of energy densities where $\overline{n_g} \simeq 0$. This can already be seen in the top panels of Figs.\ \ref{fig_low} and \ref{fig_high}; however, it is more evident if we look at the first and second derivatives of $\overline{n_g}(\varepsilon)$, reported in Fig.\ \ref{fig_dng} in the case $\alpha = 3\times 10^{-5}$.  
The peak in the second derivative of $\overline{n_g}(\varepsilon)$ can be effectively taken as the upper limit of a ``highly clustered phase'' (which is not a ``true'' phase because there is no singularity). The existence of such a region is a very interesting feature of this model: it reminds what has been observed in simulations of confined and regularized three-dimensional self-gravitating systems \cite{IspolatovCohen,DeVega}, where one finds a low-energy clustered phase, an intermediate-energy ``core-halo'' phase (which would be mimicked by the coexistence of gas and cluster particles in the effective model) and a high-energy perfect gas phase. We shall come back to this point in the two final sections of the paper. 

An unphysical feature of the model is that when $\varepsilon$ is very large the number of gas particles starts to decrease and eventually $\overline{n_g}\to 0$ for $\varepsilon \to \infty$. This happens for any value of $\alpha$ and is due to the fact that the coordinates of the cluster particles are allowed to take values in $\mathbb{R}$ instead of in $(-\pi,\pi]$, so that at very high energy it is always convenient to make a ``loose'' cluster whose effective size is larger than the circle. However, this happens at values of $\varepsilon$ which get larger and larger as $\alpha$ gets smaller, so that one can safely ignore this fact in practice: in Fig.\ \ref{fig_low} one can see a first hint of this phenomenon, which is instead invisible in the energy range of Fig.\ \ref{fig_high}. 

A feature of the model that does not compare well with the SGR model is that for any value of $\alpha$ there is a region of nonconcave entropy: there is always ensemble inequivalence, so that, as we shall see in the following, there is no canonical tricritical point and the phase transition in the canonical ensemble is always discontinuous. In the SGR model, on the contrary, the ensemble inequivalence is present only for $\alpha < \alpha_{\mathrm{CT}}$ with $\alpha_{\mathrm{CT}} \simeq 0.1$.  On the other hand, the approximations made to derive the effective model are only reasonable for small values of $\alpha$, so that this is not a big surprise.

\subsection{Canonical ensemble}

Let us now discuss the solution of the effective model in the canonical ensemble. We have to compute the partition function: the calculation is even more straightforward than in the microcanonical case, so that we report the details of the calculation in \ref{app_can}. It turns out that in the thermodynamic limit $N\to \infty$ the free energy density is given by
\beq
f(\beta) = \inf_{n_g\in [0,1]} f(\beta,n_g)
\label{fmin}
\eeq
where $\beta = T^{-1}$ (we set the Boltzmann cosntant to unity) and
\begin{eqnarray}
f(\beta,n_g) & = & \frac{n_g}{\beta}\log n_g + \frac{1-n_g}{\beta} \log(1-n_g) - \frac{n_g}{\beta} \log(2\pi) - \frac{1}{2\beta} \log \left(\frac{2\pi}{\beta} \right) \nonumber \\
& - & \frac{n_g \gamma(n_g -2)}{2\sqrt{2}} - \frac{(1 - n_g)^2}{2\sqrt{2\alpha}} - \frac{1 - n_g}{2\beta} \log\left[\frac{2\pi (2\alpha)^{3/2}}{\beta(1 - n_g)} \right]~. \label{f(ng)}
\end{eqnarray}
Again, the solution amounts to finding the value $\overline{n_g}(\beta)$ of $n_g$ realizing the extremum in (\ref{fmin}). This can be easily done numerically, since the explicit form (\ref{f(ng)}) of $f(\beta,n_g)$ is known. 

\subsubsection{Results for the thermodynamic quantities}

\begin{figure}
\center
\psfrag{T}{$T$}
\psfrag{e}{$u$}
\psfrag{0.5}{0.5}
\psfrag{1}{1}
\psfrag{1.5}{1.5}
\psfrag{2}{2}
\psfrag{2.5}{2.5}
\psfrag{3}{3}
\psfrag{1}{1}
\psfrag{-1}{-1}
\psfrag{-2}{-2}
\psfrag{-3}{-3}
\includegraphics[width=12cm,clip=true]{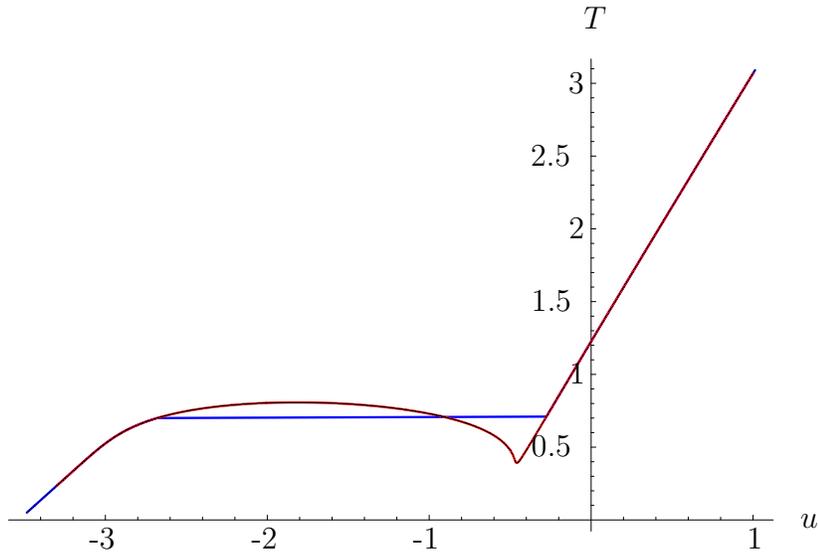}
\caption{Effective model, $\alpha = 10^{-2}$. Comparison between the caloric curve $T(u)$ computed in the canonical ensemble (blue line) and that computed in the microcanonical ensemble (red  line; already reported in the bottom panel of Fig.\ \protect\ref{fig_low}). On the horizontal axis there is $u$ in the canonical case and $\varepsilon$ in the microcanonical case.}
\label{fig_can_low}
\end{figure}

In Fig.\ \ref{fig_can_low} we report the results for the energy density $u = \frac{d(\beta f)}{d\beta}$, i.e.,
\beq
u(\beta) = \frac{1}{2\beta} + \frac{1 - \overline{n_g}}{2\beta} -  \frac{\overline{n_g} \gamma(2- \overline{n_g})}{2\sqrt{2}} - \frac{\left(1 - \overline{n_g}\right)^2}{2\sqrt{2\alpha}}~,
\eeq
for $\alpha = 10^{-2}$ and we compare them with the microcanonical results.
In order to make the comparison easier, we do not plot $u(\beta)$ or $u(T)$; we rather report $T$ as a function of $u$, i.e., the canonical caloric curve. 

As already noted, there is ensemble inequivalence for any value of the softening parameter $\alpha$. Moreover, the region of inequivalence is larger than in the SGR model: for instance, when $\alpha = 10^{-2}$ the inequivalence region is bounded below by an energy density $u_{\mathrm{low}} \simeq -3.15$, while in the SGR model the bound is at $u_{\mathrm{low}} \simeq -1.98$. 

\section{Thirring model}

We now discuss the relation between our model and a model introduced by W.\ Thirring forty years ago \cite{Thirring} as a toy model of a system\footnote{Thirring called it ``A somewhat artificial version of a star."} with negative specific heat. The latter model is solvable up to a maximization in a single variable, as our effective model. With a suitable choice of the parameters the microcanonical thermodynamics of the Thirring model is qualitatively similar to that of self-gravitating systems, so that we expect it to be close to that of the effective model introduced above. It will turn out that the two models are even more closely related than one would {\em a priori} suspect.

To define the Thirring model, consider $N$ classical particles confined in a three-dimensional volume $V$ and a volume $V_0 \subset V$, which will play the r\^{o}le of the cluster. The Hamiltonian\footnote{Note that, at variance with 
Ref.\ \protect\cite{Thirring}, we used the Kac prescription to make the Hamiltonian extensive. This will allow us to study the model in the ``conventional'' thermodynamic limit.} is
\beq
{\cal H} = \frac{1}{2}\sum_{i = 1}^N \mathbf{p}_i^2 + \frac{1}{2N} \sum_{i,j = 1}^N v(\mathbf{r}_i,\mathbf{r}_j) 
\eeq
where the potential energy is highly nonlocal, i.e., $v$ is such that particles $i$ and $j$ interact with constant energy if they are both in $V_0$ and are free otherwise:
\beq
v(\mathbf{r}_i,\mathbf{r}_j) = -2 \nu \chi_{V_0}(\mathbf{r}_i)\chi_{V_0}(\mathbf{r}_j)~,   
\eeq
where $\nu > 0$ (the force is attractive) and $\chi_{V_0}$ is the characteristic function of the subset $V_0$, i.e., $\chi_{V_0}(x) = 1$ if $x\in V_0$ and $\chi_{V_0}(x) = 0$ if $x\not\in V_0$. Following \cite{Thirring}, we write the volume outside $V_0$ as
\beq
V - V_0 = e^F V_0\,\, ;
\eeq
when $e^F\gg 1$, i.e., the volume outside $V_0$ is much larger than $V_0$, in the microcanonical ensemble the model is solvable in the thermodynamic limit $N\to \infty$ up to a maximization on a single variable, the number $N_0$ of particles in $V_0$. Such a maximization comes from the evaluation of the density of states containing all the possible values of $N_0$ via a saddle-point approximation, similarly to what we have done for the effective model. Details on the solution can be found in \cite{Thirring}. The entropy density can be written 
\beq
s(\varepsilon) = \sup_{n_0 \in [n_0^{\mathrm{min}},1]} s(\varepsilon,n_0) 
\eeq 
where $\varepsilon = \frac{E}{N}$ is the energy density, $n_0 = \frac{N_0}{N}$ is the fraction of particles in $V_0$, $n_0^{\mathrm{min}}$ is the minimum value of $n_0$ allowed at a given $\varepsilon$ and 
\beq
s(\varepsilon,n_0) = \frac{3}{2}\log\left(\frac{2\pi\varepsilon}{3} \right) + \log\left[\frac{(\varepsilon + \nu n_0^2)^{3/2} e^{F(1-n_0) +1}}{n_0^{n_0}(1 - n_0)^{1-n_0}} \right]\,\, , \label{s(n0)}
\eeq 
up to an irrelevant additive constant. In the following we shall denote by $\overline{n_0}$ the value of $n_0$ that maximizes the entropy (\ref{s(n0)}).

\begin{figure}
\center
\psfrag{k}{$1-\overline{n_0}$}
\psfrag{e}{$\varepsilon$}
\psfrag{1}{1}
\psfrag{0.8}{0.8}
\psfrag{0.6}{0.6}
\psfrag{0.4}{0.4}
\psfrag{0.2}{0.2}
\psfrag{1}{1}
\psfrag{-0.2}{-0.2}
\psfrag{-0.4}{-0.4}
\psfrag{-0.6}{-0.6}
\psfrag{-0.8}{-0.8}
\includegraphics[width=12cm,clip=true]{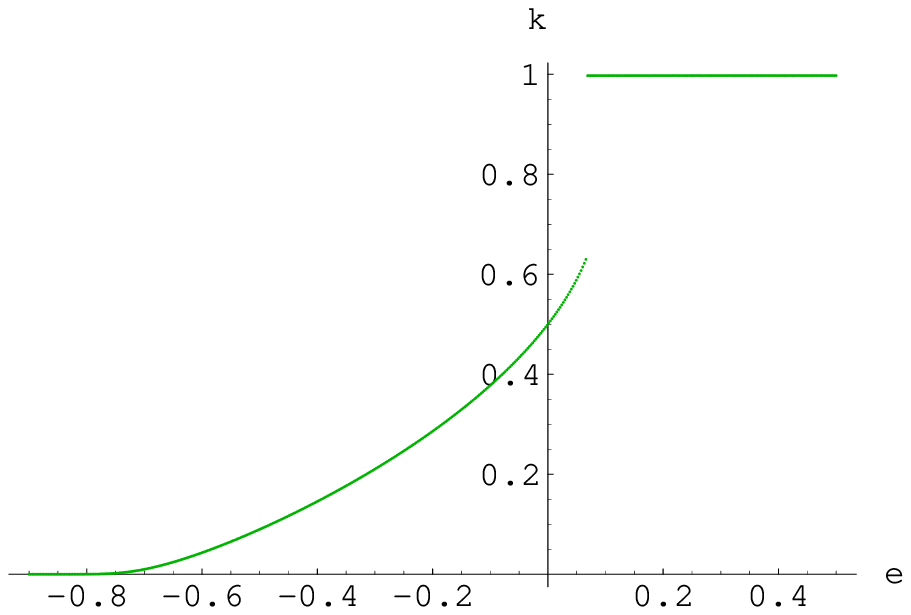}
\vspace{0.1cm}
\psfrag{T}{$T$}
\psfrag{e}{$\varepsilon$}
\psfrag{0.3}{0.3}
\psfrag{0.25}{0.25}
\psfrag{0.2}{0.2}
\psfrag{0.15}{0.15}
\psfrag{0.1}{0.1}
\psfrag{0.05}{0.05}
\psfrag{1}{1}
\psfrag{-0.2}{-0.2}
\psfrag{-0.4}{-0.4}
\psfrag{-0.6}{-0.6}
\psfrag{-0.8}{-0.8}
\includegraphics[width=12cm,clip=true]{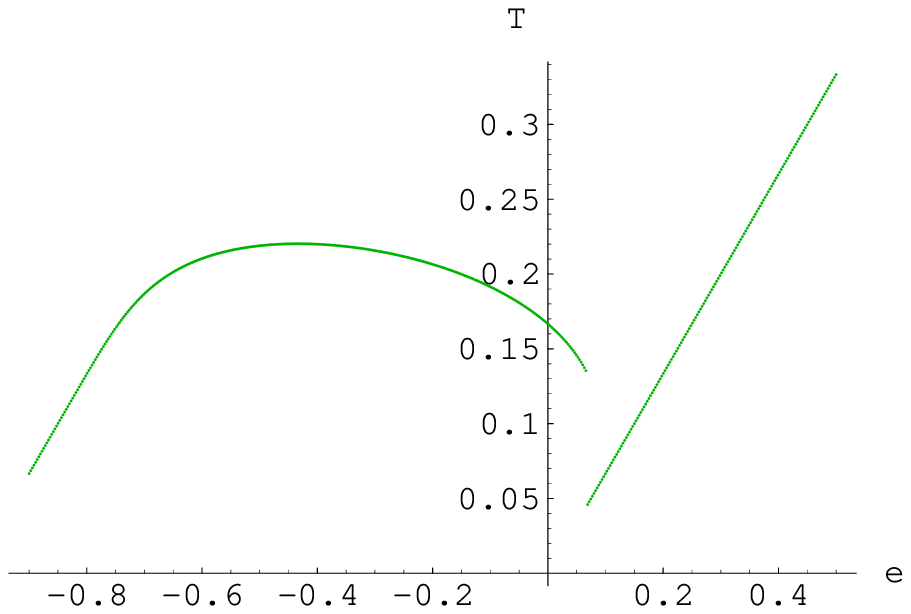}
\caption{Thirring model in the microcanonical ensemble, with $\nu=1$ and $F = 6$. (top) Fraction of particles outside $V_0$, $1 - \overline{n_0}(\varepsilon)$; (bottom) temperature $T(\varepsilon)$.}
\label{fig_thirring}
\end{figure}
Figure \ref{fig_thirring} shows an example of the thermodynamics of the Thirring model; we report $1-\overline{n_0}$ (to make the comparison with the effective model easier) and $T$ as a function of $\varepsilon$. Apart from the obvious differences in the slopes of the low- and high-energy parts of the $T(\varepsilon)$ curve, due to the different dimensionality and to the different regularization (free particles in a box {\em vs}.\ harmonic forces), the caloric curve $T(\varepsilon)$ of the Thirring model is remarkably similar to that of the effective model. This overall similarity is not a big surprise: it is a confirmation that the qualitative behavior of a confined, regularized self-gravitating system can be captured by a ``cluster + gas'' model regardless of the details on the regularization. These models are in a way the minimal models of the equilibrium statistical mechanics of self-gravitating systems. 
\begin{figure}
\center
\psfrag{a}{$-\frac{d\overline{n_0}}{d\varepsilon}$}
\psfrag{e}{$\varepsilon$}
\psfrag{2}{2}
\psfrag{1.5}{1.5}
\psfrag{1}{1}
\psfrag{0.5}{0.5}
\psfrag{0.2}{0.2}
\psfrag{0.4}{0.4}
\psfrag{-0.2}{-0.2}
\psfrag{-0.4}{-0.4}
\psfrag{-0.6}{-0.6}
\psfrag{-0.8}{-0.8}
\includegraphics[width=12cm,clip=true]{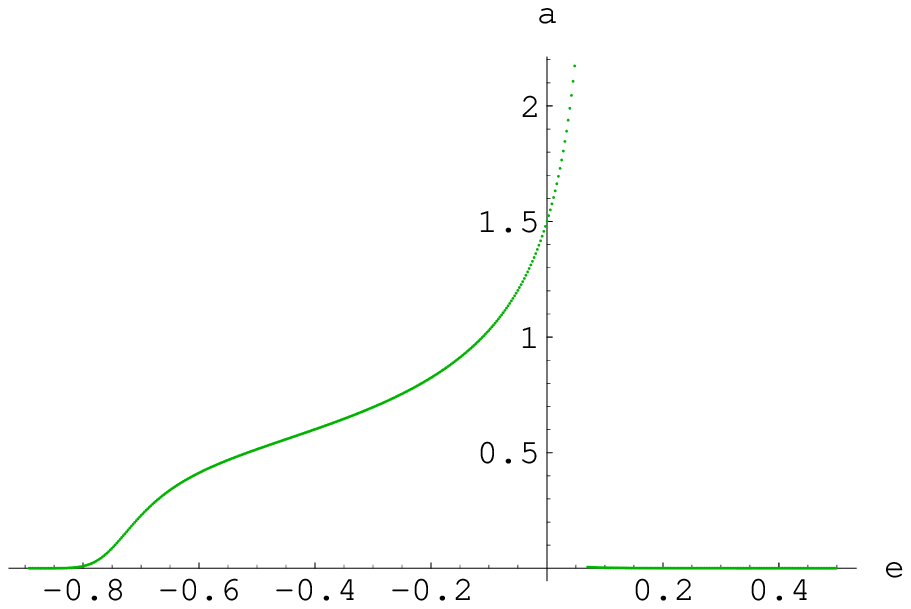}
\vspace{0.1cm}
\psfrag{b}{-$\frac{d^2\overline{n_0}}{d\varepsilon^2}$}
\psfrag{e}{$\varepsilon$}
\psfrag{6}{6}
\psfrag{4}{4}
\psfrag{2}{2}
\psfrag{-2}{-2}
\psfrag{-4}{-4}
\psfrag{0.2}{0.2}
\psfrag{0.4}{0.4}
\psfrag{-0.2}{-0.2}
\psfrag{-0.4}{-0.4}
\psfrag{-0.6}{-0.6}
\psfrag{-0.8}{-0.8}
\includegraphics[width=12cm,clip=true]{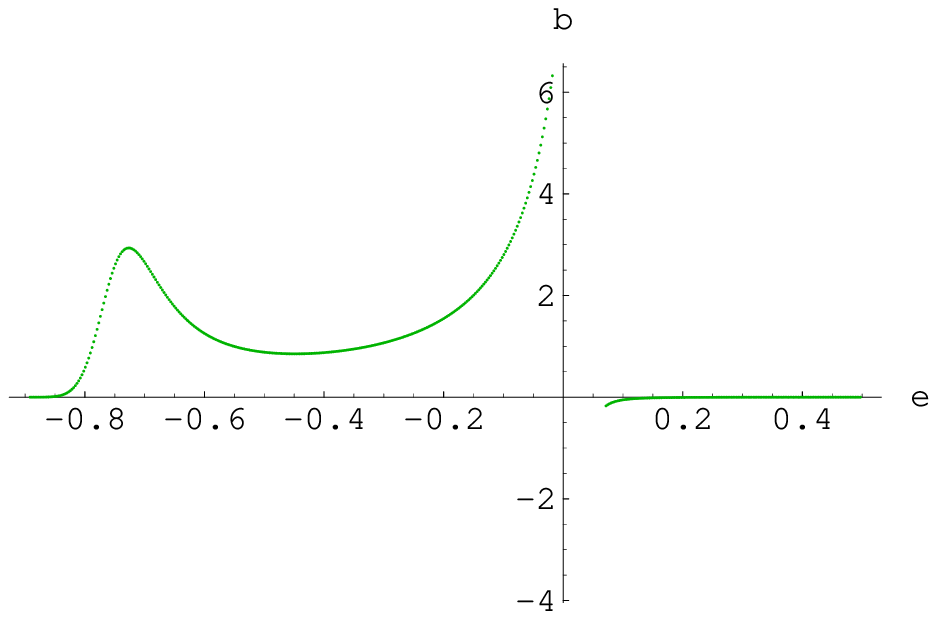}
\caption{Thirring model in the microcanonical ensemble; numerical values of the parameters as in Fig.\ \protect\ref{fig_thirring}. (top) First derivative of the fraction of particles outside $V_0$,  $-\frac{d\overline{n_0}}{d\varepsilon}$; (bottom) Second derivative $-\frac{d^2\overline{n_0}}{d\varepsilon^2}$.}
\label{fig_n0_thirring}
\end{figure}
What is more interesting is that this similarity extends to the low-energy behaviour, where one would expect a stronger dependence on the details of the regularization. Despite the fact that the small-distance regularization is completely different in the two models, Fig.\ \ref{fig_n0_thirring} shows that the behaviour of the first and second derivatives of $1-\overline{n_0}(\varepsilon)$ in the Thirring model is strikingly similar to that of $\overline{n_g}(\varepsilon)$ in the effective model (Fig.\ \ref{fig_dng}), with a strong peak in the second derivative marking the upper limit of the strongly clustered regime, where $\overline{n_0}\simeq 1$. As in the effective model, due to degeneracy factor in the density of states there are no phases with $\overline{n_0}\equiv 1$ or $\overline{n_0}\equiv 0$; at high energies, in the gas phase,  $\overline{n_0}$ is very small but always strictly positive\footnote{Due to the fact that the cluster size is fixed the Thirring model does not show the unphysical decrease of the fraction of particles outside the cluster at very high energies.}.

At variance with the Thirring model, the effective model we have introduced above is derived from a ``microscopic'' model, the SGR model. This feature allows a detailed comparison between the thermodynamics of the effective model and that of the ``microscopic'' one, allowing to test the effect of the approximations. Apart from this feature, our effective model is very similar to the Thirring model, and can be considered as a way to derive ``Thirring-like'' (i.e., ``cluster + gas'') toy models from models with true gravitational interactions.

\section{Conclusions and outlook}

We have discussed how to derive an effective model of a self-gravitating system starting from the SGR model introduced and studied in Refs.\ \cite{SGR1,SGR2}. Such an effective model is solvable up to a numerical maximization in a single variable (which plays the r\^ole of an order parameter) in both the microcanonical and the canonical ensembles. Following a suggestion coming from the study of the dynamics of the SGR model \cite{SGR1,SGR2}, the effective model assumes that particles can be split into two families: cluster particles, all of which are interacting with each other by harmonic forces, and gas particles, which feel only a constant potential due to both the cluster particles and the mean field of the other gas particles. The fraction of gas particles, $n_g$,  is the ``order parameter'' to be determined self-consistently.

Despite the rather crude approximations used to derive the effective model, the results for the thermodynamic quantities are quite close to those found for the SGR model using the numerical method developed in \cite{SGR2}, especially for small values of the softening parameter $\alpha$, where the behaviour of the two systems is closer to that of an ``ideal'' self-gravitating system. Although one could expect an agreement at small energies, the agreement is definitely good also at energies up to and above the transition between the phase with negative specific heat and the homogeneous phase, which is well reproduced by our results. It is interesting that such a simple toy model as ours is able to reproduce the thermodynamics of the SGR model even at quantitative level.

A qualitative disagreement appears at larger values of the softening parameter, where in the SGR model the entropy becomes concave and there is no longer ensemble inequivalence: in our effective model the entropy appears to be always nonconcave, for any value of $\alpha$; however, the approximations made are no longer justified when $\alpha$ is not small. 

Clearly, there is room for improvements: the caloric curve $T(\varepsilon)$ is well reproduced when $\alpha$ is small, but its shape in the low-energy part---see e.g.\ Fig.\ \ref{fig_high}---shows that there are important anharmonic effects which should be taken into account. 

We have also compared our model with another model introduced by Thirring in 1970 \cite{Thirring}. In that model one has a cluster of particles confined in a finite region of space, all interacting with each other via a constant attractive potential. This region of space is enclosed in larger volume where particles can escape becoming free particles. When the cluster extension is small, the thermodynamics of the two models is very similar, also in the low energy region dominated by the regularization of the potential, although the nature of the cluster is different in the two models. This shows, in our opinion, that ``cluster + gas'' toy models like these are good candidates as minimal models of self-gravitating systems. 

An interesting feature of our model (and of the Thirring model too, although this had not been noticed before, to the best of our knowledge) is the presence of a low-energy region where the fraction of gas particles $n_g$ is very small and stays very small up to a certain energy where it starts rising rapidly. There is a mathematical reason why $n_g$ can not be exactly zero in a finite region of energy, i.e., the degeneracy factor associated with counting the number of ways of constructing a state with a given fraction of gas particles. This counting assumes that all the gas particles are independent, which is clearly an oversimplification. One could wonder whether a more refined counting may imply the presence of a singularity at low energy, bounding a phase with $n_g\equiv 0$, whose existence for the SGR model has been conjectured in \cite{gravring}. A hint in this direction comes from assuming that the gas particles are all correlated like bosons (which is wrong, but is somehow the opposite situation to that considered here), so that states obtained from each other by interchanging two gas particles should not be considered as distinct and no degeneracy factor would be present in the density of states. Such a calculation has been performed in \cite{cesare} and yields a sharp transition at a finite energy density and a phase with $n_g\equiv 0$. The drawback is that the agreement with the thermodynamics of the SGR model is definitely worse. It is tempting to think that maybe a weaker transition occurs, and that it can be described by a ``proper'' counting of the degeneracy of the gas particle states.

\section*{Acknowledgments}

We thank S.\ Ruffo for very useful discussions on the subject of the present paper.

\appendix

\section{Density of states and entropy}
\label{app_micro}

To solve the model in the microcanonical ensemble we need to calculate the density of states (\ref{omegaeff}), i.e., 
\begin{eqnarray}
{\omega}_{N} (E) & = &  \sum_{N_g = 0}^N \frac{N!}{N_g!\left(N - N_g \right)!} \int_{-\infty}^{\infty} dp_{1} \cdots dp_{N}\int_{-\pi}^\pi d\vartheta_{1} \cdots d\vartheta_{N_g}  \int_{-\infty}^\infty d\vartheta_{N_g+1} \cdots d\vartheta_{N}\,\nonumber \\
& \times & \delta \left[\frac{1}{2}\sum_{i = 1}^N p_i^2 + \frac{\mu}{2}\sum_{i,j=N_g + 1}^{N}\left(\vartheta_i -  \vartheta_j \right)^2 - V_0 -{E}\right] ~. \label{omegaeff_app} 
\end{eqnarray}
To compute the above integral we follow a procedure similar to that used in \cite{KastnerSchnetz,CasettiKastner}. First, we expand the square in the last sum, obtaining
\begin{eqnarray}
{\omega}_{N} (E) & = & \frac{1}{\mu}\sum_{N_g = 0}^N \frac{N!}{N_g!\left(N - N_g \right)!} \int_{-\infty}^{\infty} dp_{1} \cdots dp_{N}\int_{-\pi}^\pi d\vartheta_{1} \cdots d\vartheta_{N_g} \int_{-\infty}^\infty d\vartheta_{N_g+1} \cdots d\vartheta_{N}\,  \nonumber \\
& \times & \delta \left[ \frac{1}{2\mu}\sum_{i = 1}^N p_i^2 + (N-N_g)\sum_{i=N_g + 1}^{N} \vartheta_i^2 - \left(\sum_{i=N_g + 1}^{N} \vartheta_i \right)^2 - \frac{V_0 +E}{\mu}\right] ~, \label{omegaeff2} 
\end{eqnarray}
and then we search for a coordinate transformation diagonalizing the coupling between the $\vartheta$'s. As to this point, we note that
\beq
(N-N_g)\sum_{i=N_g + 1}^{N} \vartheta_i^2 - \left(\sum_{i=N_g + 1}^{N} \vartheta_i \right)^2 = 
(\vartheta_{N_g + 1},\ldots,\vartheta_{N}) \mathbb{A} (\vartheta_{N_g + 1},\ldots,\vartheta_{N})^T~, 
\eeq
where the symmetric $(N-N_g) \times (N-N_g)$ matrix $\mathbb{A}$ reads as
\beq
\mathbb{A} = - \left(
\begin{array}{ccc}
1 & \cdots & 1\\
\vdots & \ddots & \vdots \\
1 & \cdots & 1
\end{array} 
 \right)
 + (N-N_g)\mathbb{I}_{N-N_g}
\eeq
and $\mathbb{I}_{d}$ is the $d \times d$ identity matrix. The matrix $\mathbb{A}$ has eigenvalues $\lambda_1 = 0$ and $\lambda_2 = \cdots = \lambda_{N-N_g} = N-N_g$, and can be diagonalized by an orthogonal transformation that does not change the integration measure. Hence we can write
\begin{eqnarray}
{\omega}_{N} (E) & = & \frac{1}{\mu}\sum_{N_g = 0}^N \frac{N!}{N_g!\left(N - N_g \right)!} (2\pi)^{N_g} \int_{-\infty}^{\infty} dp_{1} \cdots dp_{N} \int_{-\infty}^\infty d\vartheta_{N_g +1} \cdots d\vartheta_{N}\,  \nonumber \\
& \times & \delta \left[ \frac{1}{2\mu}\sum_{i = 1}^N p_i^2 + (N-N_g)\sum_{i=N_g + 1}^{N-1} \vartheta_i^2 - \frac{V_0 +E}{\mu}\right] ~, \label{omegaeffdiag} 
\end{eqnarray}
where we have also performed the $N_g$ integrals over the circle. With the change of variables $\psi_i = \vartheta_i\sqrt{N-N_g}$, $i = N_g + 1, \ldots,N-1$ we get
\begin{eqnarray}
{\omega}_{N} (E) & = & \frac{1}{\mu}\sum_{N_g = 0}^N (2\pi)^{N_g} \frac{N!\left(N-N_g\right)^{-(N-N_g-1)/2}}{N_g!\left(N - N_g \right)!} \int_{-\infty}^{\infty} dp_{1} \cdots dp_{N} \int_{-\infty}^\infty d\psi_{N_g+1} \cdots d\psi_{N-1} \,  \nonumber\\
& \times & \delta \left[ \frac{1}{2\mu}\sum_{i = 1}^N p_i^2 + (N-N_g)\sum_{i=N_g + 1}^{N-1} \psi_i^2 - \frac{V_0 +E}{\mu}\right] \int_{-\infty}^\infty d\vartheta_{N}~. \label{omegaeffdiag2} 
\end{eqnarray}
The last integral in Eq.\ (\ref{omegaeffdiag2}) stems from the zero mode due to the $O(2)$ invariance of the Hamiltonian; it is divergent but does not affect the thermodynamic quantities so that from now on we will ignore it.

When $\frac{1}{2\mu}\left[2(V_0 +E) - \sum_{i = 1}^N p_i^2\right] \geq 0$, the integrals over the $\psi$'s give the volume of the $(N - N_g - 2)$-dimensional sphere $\mathbb{S}^{N - N_g - 2}_R$ of radius $R = \frac{1}{2\mu}\left[2(V_0 +E) - \sum_{i = 1}^N p_i^2\right] $; on the other hand, when $\frac{1}{2\mu}\left[2(V_0 +E) - \sum_{i = 1}^N p_i^2\right]  < 0$ the same integrals vanish. As to the integrations over the momenta, we note that the integrand depends only on 
\beq
p = \sqrt{\sum_{i = 1}^N p_i^2} ~,
\eeq
so that we can write
\begin{eqnarray}
{\omega}_{N} (E) & = & \frac{1}{\mu}\sum_{N_g = 0}^N (2\pi)^{N_g} \frac{N!\left(N-N_g\right)^{-(N-N_g-1)/2}}{N_g!\left(N - N_g \right)!} 
\frac{2\pi^{(N-N_g-1)/2}}{\Gamma\left(\frac{N-N_g-1}{2} \right)} \frac{2\pi^{N/2}}{\Gamma\left(\frac{N}{2} \right)} \,  \nonumber\\
& \times & 2^{N/2} \int_{0}^\infty dp\, p^{N-1} \left[ \frac{2E'-p^2 }{2\mu}\right]^{(N-N_g -2)/2}\Theta\left[\frac{2E'-p^2}{2\mu}\right]  ~, \label{omegaone} 
\end{eqnarray}
where $\Gamma(x)$ is the Euler gamma function, $\Theta(x)$ is the Heaviside step function and $E' = V_0 +E$. 

We want to compute 
\beq
s(\varepsilon) = \lim_{N\to\infty} \frac{1}{N} \log \omega_N(N\varepsilon)~, 
\label{sdef_app}
\eeq
where we have introduced the energy density $\varepsilon =  \frac{E}{N}$. Clearly,
\beq
\omega_N(N\varepsilon) \equiv 0 \,\, \mathrm{if}\, \, \varepsilon < \varepsilon_{\mathrm{min}}\,,
\eeq
where $\varepsilon_{\mathrm{min}}$ is the absolute minimum of the potential energy per degree of freedom,
\beq
\varepsilon_{\mathrm{min}} = -\frac{1}{2\sqrt{2\alpha}}\,\,, \label{emin}
\eeq
so that the domain of the entropy density (\ref{sdef_app}) is $\varepsilon > \varepsilon_{\mathrm{min}}$. For large $N$, computing the integral in Eq.\ (\ref{omegaone}) with the Laplace (or saddle-point) method \cite{Laplace} we get 
\begin{eqnarray}
{\omega}_{N} (N\varepsilon) & = & \frac{2^{N/2}}{\mu}\sum_{N_g = 0}^{N_g^{\mathrm{max}}(\varepsilon)} (2\pi)^{N_g} \frac{N!\left(N-N_g\right)^{-(N-N_g-1)/2}}{N_g!\left(N - N_g \right)!} 
\frac{2\pi^{(N-N_g-1)/2}}{\Gamma\left(\frac{N-N_g-1}{2} \right)} \frac{2\pi^{N/2}}{\Gamma\left(\frac{N}{2} \right)} \nonumber\\
& \times & \left[N^2 2(2\alpha)^{3/2} \right]^{(N-N_g-2)/2}\, 
\exp\left\{ N\left[ \frac{1}{2}\log\left(\frac{N}{2-n_g} \right) \right. \right. \label{omegasaddle} \\ 
& + & \left. \left. \frac{2 - n_g}{2} \log a(n_g,\alpha,\varepsilon) + \frac{1-n_g}{2} \log\left(\frac{1 - n_g}{2-n_g} \right) 
\right]\right\} + o\left(e^N \right) ~, \nonumber
\end{eqnarray}
where 
\beq
a(n_g,\alpha,\varepsilon) = \frac{\gamma}{2\sqrt{2}} n_g(2-n_g) + \frac{\left(1-n_g\right)^2}{2\sqrt{2\alpha}} + \varepsilon~, \label{a}
\eeq
and we have introduced the fraction of gas particles $n_g =  \frac{N_g}{N}$; $N_g^{\mathrm{max}}(\varepsilon)$ is the maximum number of gas particles allowed at a given energy density $\varepsilon$, so that the quantity $a(n_g,\alpha,\varepsilon)$ given by Eq.\ (\ref{a}) is positive in the domain $\varepsilon >  \varepsilon_{\mathrm{min}}$ with $0 \leq N_g \leq N_g^{\mathrm{max}}$. 

Neglecting the sub-exponential terms and using the Stirling approximation, Eq.\ (\ref{omegasaddle}) can be written as
\beq
{\omega}_{N} (N\varepsilon) = \int_0^{n_g^{\mathrm{max}}(\varepsilon)} dn_g \exp\left[N s(\varepsilon,n_g) \right]~,
\eeq
where\footnote{Eq. (\protect\ref{ngmax_app}) holds for sufficiently small $\alpha$, i.e., such that $\sqrt{2} -\gamma\sqrt{\alpha}>0$.} 
\beq
{n_g^{\mathrm{max}}(\varepsilon)} = \frac{{N_g^{\mathrm{max}}(\varepsilon)}}{N} = 1 - \sqrt{1 - \frac{1 + 2\varepsilon\sqrt{2\alpha}}{\sqrt{2} -\gamma\sqrt{\alpha}}} \label{ngmax_app}
\eeq
and
\begin{eqnarray}
s(\varepsilon,n_g) & = &\frac{1-n_g}{2}\log\left[\frac{4\pi(2\alpha)^{3/2}}{(1-n_g)(2-n_g)} \right] + \frac{1}{2} \log \left( \frac{2\pi\sqrt{2}}{2-n_g}\right) \nonumber \\
& + & \frac{2-n_g}{2} \left[1 + \log a(n_g,\alpha,\varepsilon) \right] + n_g \log (2\pi) \label{s(ng)_app}\\
& - & n_g \log n_g - (1 - n_g) \log (1 - n_g) ~; \nonumber
\end{eqnarray}
hence, we can write the entropy (\ref{sdef_app}) as
\beq
s(\varepsilon) = \sup_{n_g \in [0,n_g^{\mathrm{max}}]} s(\varepsilon,n_g) ~.
\label{smax_app}
\eeq

\section{Partition function and free energy}
\label{app_can}

We want to compute the partition function of the effective model. As in the microcanonical case, up to a factor that takes into account the correct dimensions and makes the free energy extensive, we can consider a dimensionless partition function:
\bea
Z(\beta) & = & \sum_{N_g = 0}^N \frac{N!}{N_g!(N - N_g)!} \int_{-\infty}^{\infty} dp_{1} \cdots dp_{N}\int_{-\pi}^\pi d\vartheta_{1} \cdots d\vartheta_{N_g} \nonumber \\
& \times &  \int_{-\infty}^\infty d\vartheta_{N_g+1} \cdots d\vartheta_{N}\,\exp \left(-\beta{\cal H}_{\mathrm{eff}}\right)\,\,.
\eea
Using the expression (\ref{Heff}) for ${\cal H}_{\mathrm{eff}}$ and performing the integrals over the momenta and the first $N_g$ angles we can write
\bea
Z(\beta) & = & \left(\frac{2\pi}{\beta} \right)^{N/2} \sum_{N_g = 0}^N \frac{N!}{N_g!(N - N_g)!} (2\pi)^{N_g} \exp\left\{ \beta \left[\gamma\frac{N_g(2N-N_g)}{2N\sqrt{2}} + \frac{(N - N_g)^2}{2N\sqrt{2\alpha}}  \right] \right\} \nonumber \\ 
& \times &  \int_{-\infty}^\infty d\vartheta_{N_g+1} \cdots d\vartheta_{N}\,\exp \left[-\frac{\eta(N-N_g)}{2}\sum_{i = N_g + 1}^N \vartheta_i^2 + \frac{\eta}{2} \left( \sum_{i = N_g + 1}^N \vartheta_i\right)^2\right]\,\,, \label{z}
\eea
where
\beq
\eta = \frac{\beta}{(2\alpha)^{3/2}N}\,\, .
\eeq
Using the Hubbard-Stratonovich formula 
\beq
\exp\left( \frac{b^2}{4a}\right) =  \sqrt{\frac{a}{\pi}} \int_{-\infty}^\infty dy \exp \left(-ay^2 + by \right)\,\, ,
\eeq
with $a=1$ and $b = \sqrt{2\eta}\sum_{i = N_g + 1}^N \vartheta_i$ and performing the integration over the $\vartheta$'s one gets
\bea
Z(\beta) & = & \left(\frac{2\pi}{\beta} \right)^{N/2} \sum_{N_g = 0}^N \frac{N!}{N_g!(N - N_g)!} (2\pi)^{N_g} \exp\left\{ \beta \left[\gamma\frac{N_g(2N-N_g)}{2N\sqrt{2}} + \frac{(N - N_g)^2}{2N\sqrt{2\alpha}}  \right] \right\} \nonumber \\ 
& \times &  \sqrt{\frac{1}{\pi}} \left[ \frac{2\pi}{\eta(N-N_g)}\right]^{(N-N_g/2)}
\int_{-\infty}^\infty dy \,\,. \label{z2}
\eea
The integral in (\ref{z2}) is divergent (as in the microcanonical case, it comes from the zero mode associated to the rotational invariance of the potential) but does not affect the thermodynamic quantities and will be discarded from now on. 

The free energy density is 
\beq
f(\beta) = -\frac{1}{\beta} \lim_{N\to\infty} \frac{1}{N}\log Z(\beta).
\eeq
In the limit $N\to \infty$, the sum over $N_g$ in (\ref{z2}) is dominated by its largest term;  using the Stirling approximation and introducing the fraction of gas particles $n_g = \frac{N_g}{N}$ we can thus write   
\beq
f(\beta) = \inf_{n_g\in [0,1]} f(\beta,n_g)
\label{fmin_app}
\eeq
where 
\begin{eqnarray}
f(\beta,n_g) & = & \frac{n_g}{\beta}\log n_g + \frac{1-n_g}{\beta} \log(1-n_g) - \frac{n_g}{\beta} \log(2\pi) - \frac{1}{2\beta} \log \left(\frac{2\pi}{\beta} \right) \nonumber \\
& - & \frac{n_g \gamma(2- n_g)}{2\sqrt{2}} - \frac{(1 - n_g)^2}{2\sqrt{2\alpha}} - \frac{1 - n_g}{2\beta} \log\left[\frac{2\pi (2\alpha)^{3/2}}{\beta(1 - n_g)} \right]~. \label{f(ng)_app}
\end{eqnarray}

\end{document}